\documentstyle[12pt]{article}
\catcode`\@=11
\@addtoreset{equation}{section}

\def\theequation{\thesection.\arabic{equation}}

\newtoks\@stequation

\def\subequations{\refstepcounter{equation}%
  \edef\@savedequation{\the\c@equation}%
  \@stequation=\expandafter{\theequation}
  \edef\@savedtheequation{\the\@stequation}
  \edef\oldtheequation{\theequation}%
  \setcounter{equation}{0}%
  \def\theequation{\oldtheequation\alph{equation}}}

\def\endsubequations{\setcounter{equation}{\@savedequation}%
  \@stequation=\expandafter{\@savedtheequation}%
  \edef\theequation{\the\@stequation}\global\@ignoretrue}

\def\hybrid{\topmargin -20pt	\oddsidemargin 0pt
	\headheight 0pt	\headsep 0pt
	\textwidth 6.25in	
	\textheight 9.5in	
	\marginparwidth .875in
	\parskip 5pt plus 1pt	\jot = 1.5ex}

\def\baselinestretch{1.2}

\catcode`\@=11

\def\marginnote#1{}
\newcount\hour
\newcount\minute
\newtoks\amorpm
\hour=\time\divide\hour by60
\minute=\time{\multiply\hour by60 \global\advance\minute by-\hour}
\edef\standardtime{{\ifnum\hour<12 \global\amorpm={am}%
	\else\global\amorpm={pm}\advance\hour by-12 \fi
	\ifnum\hour=0 \hour=12 \fi
	\number\hour:\ifnum\minute<10 0\fi\number\minute\the\amorpm}}
\edef\militarytime{\number\hour:\ifnum\minute<10 0\fi\number\minute}
\def\draftlabel#1{{\@bsphack\if@filesw {\let\thepage\relax
   \xdef\@gtempa{\write\@auxout{\string
      \newlabel{#1}{{\@currentlabel}{\thepage}}}}}\@gtempa
   \if@nobreak \ifvmode\nobreak\fi\fi\fi\@esphack}
	\gdef\@eqnlabel{#1}}
\def\@eqnlabel{}
\def\@vacuum{}
\def\draftmarginnote#1{\marginpar{\raggedright\scriptsize\tt#1}}

\def\draft{\oddsidemargin -.2truein
	\def\@oddfoot{\sl preliminary draft \hfil
	\rm\thepage\hfil\sl\today\quad\militarytime}
	\let\@evenfoot\@oddfoot	\overfullrule 3pt
	\let\label=\draftlabel
	\let\marginnote=\draftmarginnote
   \def\@eqnnum{(\theequation)\rlap{\kern\marginparsep\tt\@eqnlabel}%
\global\let\@eqnlabel\@vacuum}  }

\def\preprint{\twocolumn\sloppy\flushbottom\parindent 2em
	\leftmargini 2em\leftmarginv .5em\leftmarginvi .5em
	\oddsidemargin -.5in	\evensidemargin -.5in
	\columnsep .4in	\footheight 0pt
	\textwidth 10.in	\topmargin  -.4in
	\headheight 12pt \topskip .4in
	\textheight 6.9in \footskip 0pt
	\def\@oddhead{\thepage\hfil\addtocounter{page}{1}\thepage}
	\let\@evenhead\@oddhead	\def\@oddfoot{}	\def\@evenfoot{} }

\def\numberbysection{\@addtoreset{equation}{section}
	\def\theequation{\thesection.\arabic{equation}}}

\def\titlepage{\@restonecolfalse\if@twocolumn\@restonecoltrue\onecolumn
     \else \newpage \fi \thispagestyle{empty}\c@page\z@
	\def\thefootnote{\fnsymbol{footnote}} }

\def\endtitlepage{\if@restonecol\twocolumn \else \newpage \fi
	\def\thefootnote{\arabic{footnote}}
	\setcounter{footnote}{0}}  

\catcode`@=12
\relax

\def\figcap{\section*{Figure Captions\markboth
	{FIGURECAPTIONS}{FIGURECAPTIONS}}\list
	{Figure \arabic{enumi}:\hfill}{\settowidth\labelwidth{Figure
999:}
	\leftmargin\labelwidth
	\advance\leftmargin\labelsep\usecounter{enumi}}}
 \relax
\def\tablecap{\section*{Table Captions\markboth
	{TABLECAPTIONS}{TABLECAPTIONS}}\list
	{Table \arabic{enumi}:\hfill}{\settowidth\labelwidth{Table
999:}
	\leftmargin\labelwidth
	\advance\leftmargin\labelsep\usecounter{enumi}}}
 \relax
\def\reflist{\section*{References\markboth
	{REFLIST}{REFLIST}}\list
	{[\arabic{enumi}]\hfill}{\settowidth\labelwidth{[999]}
	\leftmargin\labelwidth
	\advance\leftmargin\labelsep\usecounter{enumi}}}
 \relax
\makeatletter
\newcounter{pubctr}
\def\publist{\@ifnextchar[{\@publist}{\@@publist}}
\def\@publist[#1]{\list
	{[\arabic{pubctr}]\hfill}{\settowidth\labelwidth{[999]}
	\leftmargin\labelwidth
	\advance\leftmargin\labelsep
	\@nmbrlisttrue\def\@listctr{pubctr}
	\setcounter{pubctr}{#1}\addtocounter{pubctr}{-1}}}
\def\@@publist{\list
	{[\arabic{pubctr}]\hfill}{\settowidth\labelwidth{[999]}
	\leftmargin\labelwidth
	\advance\leftmargin\labelsep
	\@nmbrlisttrue\def\@listctr{pubctr}}}
 \relax
\makeatother
\newskip\humongous \humongous=0pt plus 1000pt minus 1000pt

\newif\ifdtup

\relax
\hybrid
\def\s{\sigma}
\def\thefootnote{\fnsymbol{footnote}}
\def\be{\begin{equation}}
\def\ee{\end{equation}}
\def\bs{\begin{subequations}}
\def\es{\end{subequations}}
\def\ba{\begin{eqnarray}}
\def\ea{\end{eqnarray}}
\def\d{\partial}
\def\db{\bar{\partial}}
\def\G{\Gamma}

\def\Jb{\bar{J}}
\def\a{\alpha}
\def\th{\theta}

\def\Jb{\bar{J}}

\def\P{\Phi}

\def\e{\epsilon}

\def\ti{\times}
\def\bA{\bar{A}}
\def\o{\omega}
\def\ra{\rightarrow}
\def\rd{{\rm d}}
\begin{document}
\renewcommand{\theequation}{\arabic{equation}}
\newcommand{\beq}{\begin{equation}}
\newcommand{\eeq}[1]{\label{#1}\end{equation}}
\newcommand{\ber}{\begin{eqnarray}}
\newcommand{\eer}[1]{\label{#1}\end{eqnarray}}
\begin{titlepage}
\begin{center}

\hfill CERN-TH.7310/94\\
\hfill CPTH-A309.0694 \\
\hfill hep-th/9406082  \\

\vskip .3in

{\large \bf A New Duality Symmetry in String Theory}
\vskip .6in

{\bf Elias Kiritsis}
\\
\vskip .1in

{\em Theory Division, CERN, CH-1211\\
Geneva 23, SWITZERLAND} \footnote{e-mail address:
KIRITSIS@NXTH04.CERN.CH}\\
\vskip .2in
and
\vskip .2in
{\bf N.A. Obers}
\vskip .1in
{\em Centre de Physique Th\'eorique
\footnote{Laboratoire Propre du CNRS UPR A.0014}  \\
Ecole Polytechnique \\
F-91128 Palaiseau,
FRANCE } \footnote{e-mail address: OBERS@ORPHEE.POLYTECHNIQUE.FR}

\end{center}

\vskip .3in

\begin{center} {\bf ABSTRACT } \end{center}
\begin{quotation}\noindent

We consider the conformal gauging of non-abelian groups. In such
cases
there are inequivalent ways of gauging (generalizing the axial and
vector
cases for abelian groups) corresponding to external automorphisms of
the group.
Different $\s$-models obtained this way correspond to the same
conformal field theory. We use the method of quotients to formulate
this equivalence as a new duality symmetry.
\end{quotation}
\vskip 2.0cm
CERN-TH.7310/94\\
June 1994\\
\end{titlepage}
\vfill
\eject
\def\baselinestretch{1.2}
\baselineskip 16 pt
\noindent
\setcounter{equation}{0}

Duality symmetries are special to string theory and seem to be very
useful
in understanding stringy physics. Although discovered in flat
torroidal backgrounds \cite{dual1}, they were shown to persist
semiclassically
in curved backgrounds with abelian \cite{buscher} or non-abelian
\cite {que-ossa} symmetries.
In \cite{k1,dvv} it was realized that axial and vector gauging of an
abelian chiral symmetry provides with two dual versions of the same
$\s$-model.
One on the other hand can use CFT arguments in order to show that
axial and vector abelian cosets correspond to the same CFT,
\cite{k2}.
Axial-vector duality was employed in \cite{rv} to generate, using the
method of quotients, the general abelian duality transformations,
\cite{buscher}, and used
to generalize the $O(d,d,Z)$ symmetry to curved backgrounds
\cite{gr}.
The exactness of abelian duality symmetries in the compact case is
well understood \cite{k2,AG1}. In the  non-compact case, we know that
axial vector duality is exact, \cite{gk}, only for abelian cosets
possessing appropriate Weyl symmetries.
Concerning non-abelian duality \cite{gr2,VR,ss,AG2} the situation is
certainely not clear. It is not obvious if one has an exact symmetry
and tools like axial-vector duality are lacking so far.

In this note we will study a non-abelian analog of axial-vector
duality
which is generated by gauging non-abelian groups in a $\s$-model, not
only using the standard vector gauging but also other possible
gaugings related
to vector gauging by external automorphisms, \cite{bs,ao}.
In \cite{bs} it was conjectured that this will lead to another form of
duality for gauged WZW models.
As we will see, a look at the CFT construction of these gaugings
indicates
that the various cosets obtainable that way, are equivalent CFTs.
This will provide us then with some new duality transformations in
the $\s$-model picture. It is plausible that this duality (which from
now on we will label {\it quasi-axial-vector duality}) is related to
the standard non-abelian duality. Moreover in general it acts on
$\s$-model backgrounds with no isometries. An easy example of that is
$G/H$ with $H$ a maximal subgroup
of $G$.

This quasi-axial-vector duality realized partly a conjecture in
\cite{pro}.
One expects all the underlying exact symmetries of current algebra to
generate
duality symmetries for the $\s$-model description.
Affine external automorphisms however are not obvious to implement in
the $\s$-model. Here we will deal with usual external automorphisms
and provide
their implications for quasi-axial-vector duality.

The first explicit example of such duality was given in \cite{ao},
where two different gaugings were considered for the
$(E^{c}_{2}\times
E^{c}_{2})/E^{c}_{2}$ coset model. The respective $\s$-model
backgrounds were
related by a series of abelian duality transformations however.
The reason is that if one constructs the $E^{c}_{2}$ groups by
appropriately contracting $SU(2)\times U(1)$ then the
quasi-axial-vector duality is generated by standard axial-vector
duality on the U(1) before contraction.

We will start by investigating the CFT point of view.
Consider a WZW theory for some (simple) group $G$.
The standard spectrum of representations for the left and right
current algebras has the form $(R,\bar R)$, modulo the affine
truncation.
The operator product fusion rules follow again group theory modulo
truncations.
If we act on the spectrum  by an external automorphism of the right
(finite)
Lie algebra, then we obtain an equivalent theory\footnote{Sometimes
the diagonal Lie algebra may have a physical meaning. In such a case
one
can distinguish between a WZW model and its transformed version. This
happens for example in type $II_{A}$ or $II_{B}$ strings.}, in the
sense that there is a reorganization of the spectrum such that the
OPEs are the same.

A similar remark applies to coset models. Consider a non-abelian
subgroup $H$ of $G$ (such that it has non-trivial external
automorphisms).
The standard vector gauging implies the following constraint on the
spectrum
\be
J^{a}_{0}+\bar J^{a}_{0}=0
\ee
This determines the modular invariant of the $G/H$ model once the
spectrum
of the original $G$ model is known.
If we now gauge the left $H$ subgroup twisted by an automorphism $S$
we will
have a different zero mode constraint:
\be
J^{a}_{0}+{S^{a}}_{b}\bar J^{b}_{0}=0 \;\;\;\;.
\ee
The fact that the two gaugings give equivalent models is obvious once
we perform the transformation $S$ in the $G$ theory.

Once we have seen that the two gauged models correspond to the same
CFT
we can move to the $\s$-model in order to implement this equivalence
as a duality.
The strategy we use is that of quotients \cite{rv} which is tuned to
utilize
axial-vector type dualities, \cite{k1}.

The duality transformations for an abelian
isometry, may be derived by starting from a general
$(d+1)$-dimensional
$\s$-model with $U(1)_l \ti U(1)_r$ symmetry. One then has the choice
of either gauging the vector or axial $U(1)$ subgroup, leading to two
$\s$-models which are related by the abelian duality transformations.
In particular, since one can argue that both gaugings lead to the
same
CFT, this proves that the corresponding duality transformation
relates
two different backgrounds of the same CFT.

Here we will generalize this result using the
fact \cite{bs} that when we have a world-sheet action, with an $(H_l
\ti H_r)$
symmetry, we have a distinct anomaly-free
world-sheet gauging of the $H$-symmetry for
each external automorphism $S$ of $H$,
\be S_{a}{}^d S_b{}^e f_{de}{}^g (S^{-1})_g{}^c = f_{ab}{}^c
\;\;\;\;,
\;\;\;\; S_{a}{}^c S_{b}{}^d \eta_{cd} = \eta_{ab}
\;\;\;\;, \;\; a = 1, \ldots, {\rm dim}\, H
\ee
where $f_{ab}{}^c$ and $\eta_{ab}$ are the structure constants and
Killing
metric of $H$ respectively.
The statement is that the world-sheet
gauge group can be chosen to be
\be
{\cal J}_a^H = J_a^H + S_a{}^b \bar{J}_b^H
\label{wsh} \ee
where $J_a^H$ and $\bar{J}_a^H$ are the left- and right-moving world
sheet currents of $H$ respectively.
In particular, the usual vector-gauging corresponds to $S=1$, while
the validity of the axial gauging for abelian
subgroups follows since $S=-1$ is always an external automorphism in
that case.
More generally, when a non-abelian group $H$ has a non-trivial
automorphism,
the result implies that we have other gaugings beyond the vector
gauging,
which are typically of a mixed vector-axial type. For example, when
$H$ is a
simple compact
non-abelian group such automorphisms occur in $SU(n)$ and $SO(2n)$
with
$n \geq 3$, and $E_6$.

We start with the most general $\s$-model in $d+2 {\rm dim}\,H$
dimensions
that has a chiral $H_l \ti H_r$ symmetry,
\be
I(h_1,h_2,x) = I(h_1) + I(h_2)
+ \int {\rd^2 z \over \pi} [B_{ab} (x) \Jb^a_2 J_1^b + G^1_{ia}(x)
\db x^i J_1^a + G^2_{ai}(x)  \Jb_2^a \d x^i ] + I(x)
\label{act} \ee
where $I(h_I),\;I=1,2$ are two copies of the WZW action
\be
I(h) =\int_{\Sigma} {\rd^2 z \over 2 \pi } {\rm Tr}[ h^{-1} \d h
h^{-1} \db h ]
- i \int_{B} {\rd^3 z \over 6 \pi } {\rm Tr} (h^{-1} {\rm d} h)^3
\ee
satisfying the Polyakov-Wiegmann formula
\be
I(hh_0) = I(h) + I(h_0) + \int {\rd^2 z \over  \pi }{\rm Tr}[h^{-1}
\d h
\db h_0 h_0^{-1} ] \;\;\;\;.
\label{pw} \ee
The currents in the action (\ref{act}) are defined as
\be
J_I = h_I^{-1} \d h_I = J_I^a T_a \;\;\;\;, \;\;\;\;
\Jb_I = \db h_I h_I^{-1} = \Jb_I^a T_a \;\;\;\;,\;\;
I =1,2
\label{cur} \ee
where $T_a$ are matrices of some $H$-representation normalized as
${\rm Tr}(T_a T_b) = \eta_{ab}$.
Finally, the matrices $B_{ab}(x)$, $G^1_{ia}(x)$ and $G^2_{ai}(x)$
are arbitrary
functions of some set of target space coordinates $x^i,\;i=1,\ldots
,d$
with an arbitrary $\s$-model action
\be
I(x) = \int {\rd^2 z \over 2 \pi }[ E_{ij} (x) \db x^i \d x^j
+ \a' R^{(2)} \phi(x) ]
\;\;\;\;, \;\;\;\;
E_{ij} (x) = g_{ij} (x) + b_{ij}(x)
\ee
where $g_{ij}$ is the target space metric on this manifold, $b_{ij}$
the
anti-symmetric tensor and $\phi$ the dilaton field.

We remark that the form (\ref{act}) can be obtained by decomposing
the WZW
action $ I( h_1 (\th_L) g (x) h_2 (\th_R) ) $ using eq.(\ref{pw}), in
which case
a special form of the action (\ref{act}) is found\footnote{this we
can do if
the group G inside which H is embedded is large enough.}, with
\bs
\be
B_{ab} (x)= {\rm Tr} [ g(x) T_a g^{-1} (x) T_b]
\ee
\be
\db x^i G^1_{ia} (x) = {\rm Tr} [ T_a \db g(x) g^{-1} (x) ]
\;\;\;\;, \;\;\;\;
 G^2_{ai} (x) \d x^i  = {\rm Tr} [ T_a g^{-1}(x)  \d g(x) ]
 \ee
\be
I(x) = I(g(x)) \;\;\;\;.
\ee
\label{sc} \es
It is clear from the properties of the WZW action, that in this
special case
we have a chiral $H_l \ti H_r$ symmetry.
However, this symmetry persists when we drop the conditions in
eq.(\ref{sc}),
as we
will show below.

The action (\ref{act}) has  chiral $H_l \ti H_r$ symmetry, generated
by
the separately conserved left- and right-handed currents
\bs
\be
\Jb_l^a = \Jb_1^a + [ \Jb_2^c B_{cb } (x)  + \db x^i G^1_{ib}(x) ]
(\o_1)^{ba}
\ee
\be
J_r^a = J_2^a + (\o_2)^{ab} [ B_{bc} (x) J_1^c + G^2_{bi}(x) \d x^i ]
\ee
\be
\d \Jb_l^a = \db J_r^a =0
\ee
\label{cc} \es
where we have defined the matrices
\be
(\o_I)_{ab} = {\rm Tr}[ h_I T_a h_I^{-1} T_b ] \;\;\;\;, \;\; I =1,2
\label{ode} \ee
and indices are raised with the inverse Killing metric $\eta^{ab}$.

In particular, these conserved currents follow from the
chiral  transformations
\bs
\be
h_{1} \ra h_l h_ 1 \;\;\;\;, \;\;\;\;
h_l = e^{\e_l^a T_a}
\ee
\be
h_2 \ra h_2 h_r^{-1} \;\;\;\;, \;\;\;\;
h_r = e^{\e_r^a T_a}
\ee
\label{htr} \es
which, for infinitesimal $\e$, yield a variation of the action given
by
\be
\delta S = \int {\rd^2 z \over \pi} [ \d \e^a_l \Jb_a^l - \db \e^a_r
J_a^r] \;\;\;\; .
\ee
To obtain these results from (\ref{act}), one uses the
Polyakov-Wiegmann
property (\ref{pw}), along with the definition of the currents in
(\ref{cur})
and $\o_I$ in (\ref{ode}).

In the following we will assume that $H$ is such that it has at least
one
non-trivial external automorphism $S$. We can then choose to gauge
the
vector
subgroup of the $H_l \ti H_r$ symmetry or a mixed vector-axial
gauging
corresponding to a non-trivial $S$ in (\ref{wsh}). Hence, we consider
the
gauge transformations in (\ref{htr}), with
\be
\e_r^a = \e_l^b S_b{}^a
\ee
where $S=1$ is the usual vector-gauging.

The resulting gauge-invariant action is given by
\be
I_{gauge} = I(h_1,h_2,x)
 - \int {\rd^2 z \over \pi } [ A^a_l \Jb_a^l - \bA^a_r S_a {}^b J_b^r
- \bA^a_r (1-  S \o_2 B \o_1 )_{ab} A^b_l ]
\ee
where $I(h_1,h_2,x)$ as  given in (\ref{act}).
The gauged action is invariant under the left and right
$H$-transformations
in (\ref{htr}),
combined with the transformation of the gauge fields,
\be
A_l \ra h_l (A_l - \d ) h_l^{-1} \;\;\;\;, \;\;\;\; \bA_r \ra h_l
(\bA_r - \db)
h_l^{-1}
\label{gact} \ee
where the gauge field components are as usual defined by
$A_l^a = \eta^{ab} {\rm Tr}[T_b A_l]$ and
$\bA_r^a = \eta^{ab}{\rm Tr}[T_b \bA_r]$.

We will choose to gauge fix $h_2 =1 $\footnote{Notice that this is a
different gauge fixing than the one used in \cite{rv}.}, so that $J_2
= \Jb_2 =0 $ and $\o_2 =1$,
and drop the subscript 1 on the remaining quantities $h_1$, $J_1$,
$\Jb_1$
and $\o_1$.
Then, after integrating out the gauge fields in the gauged action
(\ref{gact}),
we obtain
after some algebra a $(d+{\rm dim}\,H)$-dimensional $\s$-model, whose
form
is given by
\bs
$$
I_{gauge}(h,x) = I(h) - \int {\rd^2 z \over \pi } [ G_{ab}(h,x) \Jb^a
J^b
+ \G^1_{ia}(h,x)
\db x^i J^a + \G^2_{ai}(h,x) \Jb^a \d x^i + Q_{ij}(h,x) \db x^i \d
x^j]
$$
\be
+ \int {\rd^2 z \over 2 \pi} \a' R^{(2)} \P (h,x)
 \ee
\be
J_a  = {\rm Tr} [T_a h^{-1} \d h]\;\;\;\;, \;\;\;\;
\Jb_a = {\rm Tr} [T_a \db h h^{-1} ] \;\;\;\;, \;\;\;\;
\o_{ab} = {\rm Tr} [ h T_a h^{-1} T_b ] \;\;\;\;.
\label{odef} \ee
\label{fact} \es
Here we have defined
\bs
\be
 M \equiv 1- S B \o  \;\;\;,\;\;\;
G = (1 - M^{-1} ) \o^{-1}\;\;\;,\;\;\;
\G^1  = -G^1 \o M^{-1}  \o^{-1}
\ee
\be
\G^2  = -M^{-1}S  G^2\;\;\;\,\;\;\;
Q  = -\frac{E}{2} - G^1 \o M^{-1} S G^2\;\;\;,\;\;\;
\P   = \phi  -\frac{1}{2} \ln {\rm det}\,M
\ee
\label{geo} \es
where, for brevity,  we have dropped the explicit indices of the
relevant
matrices and matrix
multiplication is employed.
The shift in the dilaton
follows from the jacobian that arises when integrating out the gauge
fields.

We remark here that due to the
the $h$-dependence of the couplings,
the model (\ref{fact}) cannot be viewed as a non-abelian
compactification.
Moreover, the $h$-dependence of the matrices $G$, $\G^{1,2}$ and $Q$
arises through $\o$ (and hence also $M$), and the $x$ dependence is
encoded
in the arbitrary matrices $B$, $G^{1,2}$ and $E$ of the parent action
(\ref{act}). So although the latter are arbitrary, the couplings in
(\ref{fact})
do have a certain form dictaded by eq.(\ref{geo}).

To summarize, starting from the parent action (\ref{act}) with some
group $H$,
we have derived for each external automorphism $S$ of $H$, a distinct
gauged
model in $d+{\rm dim}\,H$ target space dimensions. If the parent
action is
conformally invariant then so are the gauged versions of it (and vice
versa).
This implies that when the gauged theories (\ref{fact}) are
conformal, they
correspond to the same conformal field theory for any $S$.

We will now use the above to obtain quasi-axial-vector duality
transformations.
Taking the viewpoint that we are given an action of the form
(\ref{fact}) for $S=1$, we need that the quantities
\bs
\be
B(x)  = G (\o G -1)^{-1} \;\;\;,\;\;\;
G^1(x)  = \G^1 ( \o G -1 )^{-1} \;\;\;,\;\;\;
G^2 (x)  = (G\o -1 )^{-1} \G^2
\ee
\be
E(x)  = -2  Q  + 2 \G^1 \o ( G \o - 1)^{-1} \G^2 \;\;\;,\;\;\;
\phi (x)  = \Phi + \frac{1}{2} \ln {\rm det} \, (1- G\o )
\ee
\label{inv}\es
depend on $x$ only, where $\o$ is defined in (\ref{odef}). These
relations
are simply obtained by solving for $B$, $G^{1,2}$ and $E$ in
eq.(\ref{geo})
at $S=1$.

We can now substitute the expressions (\ref{inv}) into (\ref{geo})
to
obtain the corresponding dual model, given the non-trivial external
automorphism $S$ of $H$. After some manipulations we find the
quasi-axial-vector duality transformation,
\bs
\be
\tilde{G}  = S G [ 1 + \o (S-1) G ]^{-1} \;\;\;,\;\;\;
\tilde{\G}^1  = \G^1  [ 1 + \o (S-1) G ]^{-1} \;\;\;,\;\;\;
\tilde{\G}^2  = S  [ 1 + G \o (S-1)  ]^{-1} \G^2
\ee\be
\tilde{Q}  = Q - \G^1 \o (S-1)  [ 1 + G \o  (S-1)  ]^{-1}
\G^2\;\;\;,\;\;\;
\tilde{\P}  = \P - \frac{1}{2} \ln {\rm det}  [ 1 + (S-1) G \o]
\ee\label{dua}\es
relating the original background to another one, both corresponding
to the
same CFT.

As a first check on these results we note that performing the
transformations
(\ref{dua}) twice results in $S \ra S^2$, as desired by consistency.
In
particular, for $Z_2$ external automorphisms (which is the generic
case
when an
external automorphism is present) we obtain the initial background,
while
for a $Z_3$ external automorphism (which occurs for $SO(8)$) we get
back
to
the orginal background after
applying the generalized duality three times.

As another check, we choose $H$ to be abelian and evaluate
(\ref{dua}) for
$S=-1$. As our method here is the non-abelian generalization of the
one
employed in \cite{rv}, yielding the abelian duality transformations,
we should
recover those in this way.
Indeed using (\ref{dua}) in (\ref{fact}) along with
\be
I(h) = \int {\rd^2 z \over 2 \pi } \eta_{ab} \db \th^a \d \th^b
\;\;\;\;,
\;\;\; J^a = \d \th^a \;\;\;\;, \;\;\;\; \Jb^a = \db \th^a
\;\;\;\;,\;\;\;\;
\o_{ab} =\delta_{ab}
\ee
it is not difficult to establish that we obtain
the correct abelian duality transformations
\bs\be
\tilde{G}_t  = G_t^{-1} \;\;\;,\;\;\;
\tilde{\G}_t^1  = \G_t^1 G_t^{-1} \;\;\;,\;\;\;
\tilde{\G}_t^2  = -  G_t^{-1} \G_t^2
\ee\be
\tilde{Q}_t  = Q_t - \G^1_t G_t^{-1} \G^2_t \;\;\;,\;\;\;
\tilde{\P}  = \P - \frac{1}{2} \ln {\rm det} \, G_t \;\;\;\;.
\ee\label{adua}\es
Here, the total action is written as
\bs
\be
I(\theta,x) =\int {\rd^2 z \over 2 \pi } [  \db \th G_t(x) \d \th +\db
x \G^1_t
(x)
\d \th
+ \db \th \G^2_t (x)  \d x + \db x Q_t (x) \d x ]
\ee
\be
G_t = 1 - 2G\;\;\;\; , \;\;\;\; \G^1_t = -2 \G^1 \;\;\;\;, \;\;\;\;
\G^2_t = -2 \G^2 \;\;\;\; ,\;\;\;\; Q_t = - 2 Q \ee
\es
in terms of the quantities in (\ref{dua}).

\vskip .9cm

There are several questions that remain open.
The quasi-axial-vector duality symmetry presented above seems to be
in some respects different from the standard abelian axial-vector
duality.
For one thing, the backgrounds where it applies do not generically
have Killing symmetries.
Moreover a coordinate independent characterization of the backgrounds
which admit such a duality symmetry is  not obvious.
It would be interesting to study examples of dual pairs of
$\s$-models.
However, since groups with non-trivial automorphisms are rather large
the calculations are involved.
Most important, the possible connection with non-abelian duality must
be elucidated. This will help in shedding some light onto the real
differences between abelian and non-abelian duality.
It seems plausible in that respect that quasi-axial-vector dulaity is
related
to non-abelian duality upon linearizing the non-abelian symmetry.
We hope to address some of these issues in a future publication.

\noindent
{\bf Acknowledgments}

\noindent
We would like to thank L. Alvarez-Gaum\'e, I. Antoniadis,
C. Bachas,  B. Schellekens and E. Verlinde
for fruitful discussions. N.O. thanks the theory division at CERN for
hospitality and support, where part of this work was performed.

\end{document}